\documentclass[preprint,3p,twocolumn]{elsarticle}
\usepackage{graphics, color, hyperref}
\usepackage{lineno}
\usepackage{multirow}
\usepackage{subfigure}
\usepackage{array}
\usepackage{xspace}

\newcommand{\naitl}{\mbox{NaI(Tl)}\xspace}

\newcommand{\dmm}{DMM\xspace}
\newcommand{\kmm}{KMM\xspace}

\begin{document}
\journal{Astroparticle Physics}
\begin{frontmatter}

\title{Monte Carlo simulation of the SABRE PoP background}
\author[milanoinfn]{M.~Antonello} \author[melbourne]{E.~Barberio} \author[melbourne]{T.~Baroncelli} \author[princetonC]{J.~Benziger} \author[canberra]{L.~J.~Bignell} \author[milanoinfn,milano]{I.~Bolognino} \author[princeton]{F.~Calaprice} \author[lngs,gssi]{S.~Copello} \author[milanoinfn,milano]{D.~D'Angelo} \author[roma1infn]{G.~D'Imperio} \author[roma1infn]{I.~Dafinei} \author[lngs]{G.~Di Carlo} \author[roma1infn]{M.~Diemoz} \author[princeton]{A.~Di Ludovico}  \author[arc,swinburne]{A.~R.~Duffy} \author[ec]{F.~Froborg}  \author[princeton]{G.~K.~Giovanetti} \author[pnnl]{E.~Hoppe} \author[lngs]{A.~Ianni} \author[lngs]{L.~Ioannucci} \author[swinburne]{S.~Krishnan}  \author[canberra]{G.~J.~Lane} \author[melbourne]{I.~Mahmood} \author[gssi]{A.~Mariani} \author[adelaide]{P.~McGee} \author[roma1infn,roma1]{P.~Montini \tnoteref{paolo}} \author[arc,swinburne]{J.~Mould}  \author[melbourne]{F.~Nuti} \author[lngs]{D.~Orlandi}  \author[melbourne,lngs]{M.~Paris} \author[roma1infn]{V.~Pettinacci} \author[princeton]{L.~Pietrofaccia} \author[ansto]{D.~Prokopovich} \author[roma1infn,roma1]{S.~Rahatlou} \author[roma1infn]{N.~Rossi} \author[ansto]{A.~Sarbutt} \author[princeton]{E.~Shields} \author[princeton]{M.~J.~Souza} \author[canberra]{A.~E.~Stuchbery} \author[princeton]{B.~Suerfu} \author[roma1infn]{C.~Tomei} \author[melbourne]{P.~Urquijo} \author[lngs]{C.~Vignoli} \author[princeton]{M.~Wada} \author[canberra]{A.~Wallner} \author[adelaide]{A.~G.~Williams} \author[princeton]{J.~Xu} \author[melbourne]{M.~Zurowski}


\address{The SABRE Collaboration}
\address[milanoinfn]{INFN - Sezione di Milano, Milano I-20133, Italy}
\address[melbourne]{School of Physics, The University of Melbourne, Melbourne, VIC 3010, Australia}
\address[princetonC]{Chemical Engineering Department, Princeton University, Princeton, NJ 08544, USA}
\address[canberra]{Department of Nuclear Physics, The Australian National University, Canberra, ACT 2601, Australia}
\address[milano]{Dipartimento di Fisica, Universit{\`a} degli Studi di Milano, Milano I-20133, Italy}
\address[princeton]{Physics Department, Princeton University, Princeton, NJ 08544, USA}
\address[lngs]{INFN - Laboratori Nazionali del Gran Sasso, Assergi (L'Aquila) I-67100, Italy}
\address[gssi]{INFN -- Gran Sasso Science Institute, L'Aquila I-67100, Italy}
\address[roma1infn]{INFN - Sezione di Roma, Roma I-00185, Italy}
\address[arc]{ARC Centre of Excellence for All-Sky Astrophysics (CAASTRO), Australia}
\address[swinburne]{Centre for Astrophysics and Supercomputing, Swinburne University of Technology, PO Box 218, Hawthorn, Victoria 3122, Australia}
\address[ec]{Imperial College London, High Energy Physics, Blackett Laboratory, London SW7 2BZ, United Kingdom}
\address[pnnl]{Pacific Northwest National Laboratory, y, 902 Battelle Boulevard., Richland, WA 99352, USA}
\address[adelaide]{The University of Adelaide, Adelaide, South Australia, 5005 Australia}
\address[roma1]{Dipartimento di Fisica, Sapienza Universit{\`a} di Roma, Roma I-00185, Italy}
\address[ansto]{Australian Nuclear Science and Technology Organization, Lucas Heights, NSW 2234, Australia}
\tnotetext[paolo]{Currently at Dipartimento di Fisica Universit{\`a} di Roma Tre I-00146 Roma and INFN Sezione di Roma Tre - 00146, Italy}

\date{\today}

\begin{abstract}
SABRE (Sodium-iodide with Active Background REjection) is a direct
dark matter search experiment based on an array of radio-pure NaI(Tl)
crystals surrounded by a liquid scintillator veto. Twin SABRE experiments in the Northern and Southern Hemispheres will differentiate 
a dark matter signal from seasonal and local effects. The experiment is
currently in a 
Proof-of-Principle (PoP) phase, whose
goal is to demonstrate that the background rate is low enough to
carry out an independent search for a dark matter signal, with sufficient sensitivity to confirm or refute the DAMA  result during the following full-scale experimental
phase. The impact of background radiation from the detector materials
and the experimental site needs to be carefully investigated,
including both intrinsic and cosmogenically activated radioactivity. Based on the best
knowledge of the most relevant sources of background, we have
performed a detailed Monte Carlo study evaluating the expected
background in the dark matter search spectral region. The simulation
model described in this paper guides the design of the full-scale
experiment and will be fundamental for the interpretation of the
measured background and hence for the extraction of a possible dark
matter signal.
\end{abstract}

\end{frontmatter}

%
%
\section{Introduction}\label{sec:intro}
The existence of dark matter has been inferred from varied astrophysical techniques~\cite{Ade:2015xua,Conley:2011ku,Heymans:2013fya}. Several hypotheses have been formulated about the nature of dark matter, with Weakly Interacting Massive Particles (WIMPs) distributed in the galactic halo one of the most promising candidates~\cite{wimp1,wimp2}. 
A number of experiments, operated worldwide in underground laboratories, have been searching for many years for direct signals of dark matter interactions, mostly focusing on nuclear recoils produced by WIMP collisions. While most experiments with rapidly increasing sensitivity report a null observation, an observation compatible with dark matter interaction in the DAMA~\cite{DAMA/LIBRA} detector remains unverified after almost two decades. 

The dark matter signal in an Earth-based detector is expected to modulate yearly due to the change of the Earth's velocity relative to the galactic halo, produced by the orbital motion of the Earth around the Sun. The long--standing result from DAMA, an experiment comprised of a 250-kg array of highly-pure NaI(Tl) crystals at the Gran Sasso National Laboratory (LNGS), is consistent with this scenario. The modulation is observed with a robust 9.3~$\sigma$ statistical significance, and a phase compatible with the one expected for the dark matter modulation. Recently, the DAMA collaboration has released the first results from their Phase-2 experiment~\cite{DAMAPhase2}, confirming the evidence of a signal that meets all the requirements of a model-independent dark matter annual modulation signature at $12.9\,\sigma$ significance. 
When interpreted in the standard WIMP framework, the results from DAMA are in conflict with other results from experiments using different target materials~\cite{xenon1t,LUX,SuperCDMS,XMASS}. 
However such comparison is
based on several assumptions of the astrophysical and nuclear models,
and on the nature of dark matter particles and their interaction.
A model-independent test of the DAMA results is best achieved with an experiment which uses the same target material and detection technique. 
The Sodium-iodide with Active Background REjection (SABRE) experiment~\cite{SABRE,SABRETaup} is designed for this purpose and focuses on the achievement of a very low background: 
the NaI(Tl) crystals, the photosensors and all detector materials are designed for ultra-high radiopurity; in addition, active rejection of the residual background is obtained
with a liquid scintillator veto.

Presently, two arrays of NaI(Tl) crystals: the COSINE-100 experiment~\cite{COSINE2018} at the YangYang Laboratory in South Korea and the ANAIS experiment~\cite{ANAIS2017} at the Canfranc Laboratory in Spain, are in a data taking phase, with a background level 2-3 times higher than DAMA. Their results, even after several years of operation, might not resolve all possible scenarios in interpreting the DAMA signal as a dark matter signature. 

The SABRE project foresees the installation of twin detectors at LNGS (Italy) and SUPL (Stawell Underground Physics Laboratory), an underground site in Australia.
This unique combination of two high sensitivity NaI(Tl) detectors will be of great interest for dark matter research through annual modulation, beyond the goal of confirming or refuting DAMA. In fact, the dual site will provide for the first time an effective way to identify any possible season-related contribution to an observed modulation, thanks to the phase inversion between the two hemispheres.

The first phase of the SABRE experiment is the so-called Proof-of-Principle (PoP) phase and will take place at LNGS during 2018. The goal of the PoP phase is to demonstrate with a high-purity crystal operated inside a liquid scintillator veto that backgrounds are in fact low enough to carry out a reliable test of the DAMA result 
in the full-scale experiment. 

The sensitivity for suitably well-shielded experiments undertaking direct dark matter searches is generally limited by the background rate induced by radioactive contaminants in the detector material and in the materials used for the construction of the experimental setup. Such radioactive contamination may come from long-lived, naturally-occurring isotopes or from cosmogenic activation. Careful selection or development of radiopure materials and equipment is mandatory, as well as a very good knowledge of the residual radioactivity.


In this paper, we present a simulation study that evaluates the expected background of the SABRE--PoP, based on the current knowledge of the most relevant sources. 
The comparison of the simulation result with the future measurements will allow a full characterization of the detector background and provide important knowledge for the realisation of the full-scale experiment.

\section{The SABRE--PoP design and Monte Carlo simulation}\label{sec:PoPsetup}
\begin{figure}
 \centering
 \subfigure[Detector Module (internal parts, \naitl crystal in grey).]
   {\includegraphics[width=0.32\textwidth]{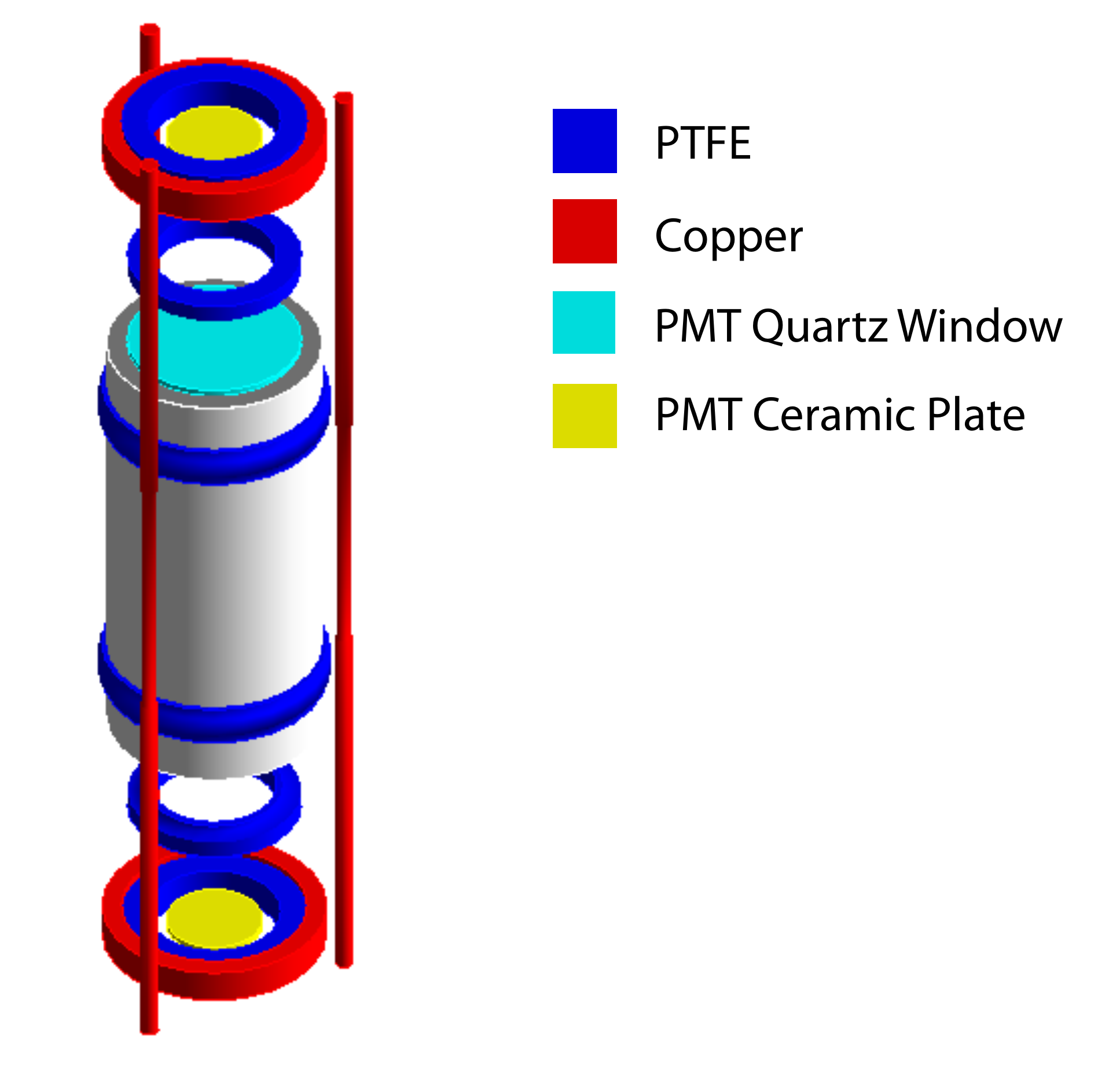}}
 \hspace{5mm}
 \subfigure[Veto PMT]
   {\includegraphics[width=0.2\textwidth]{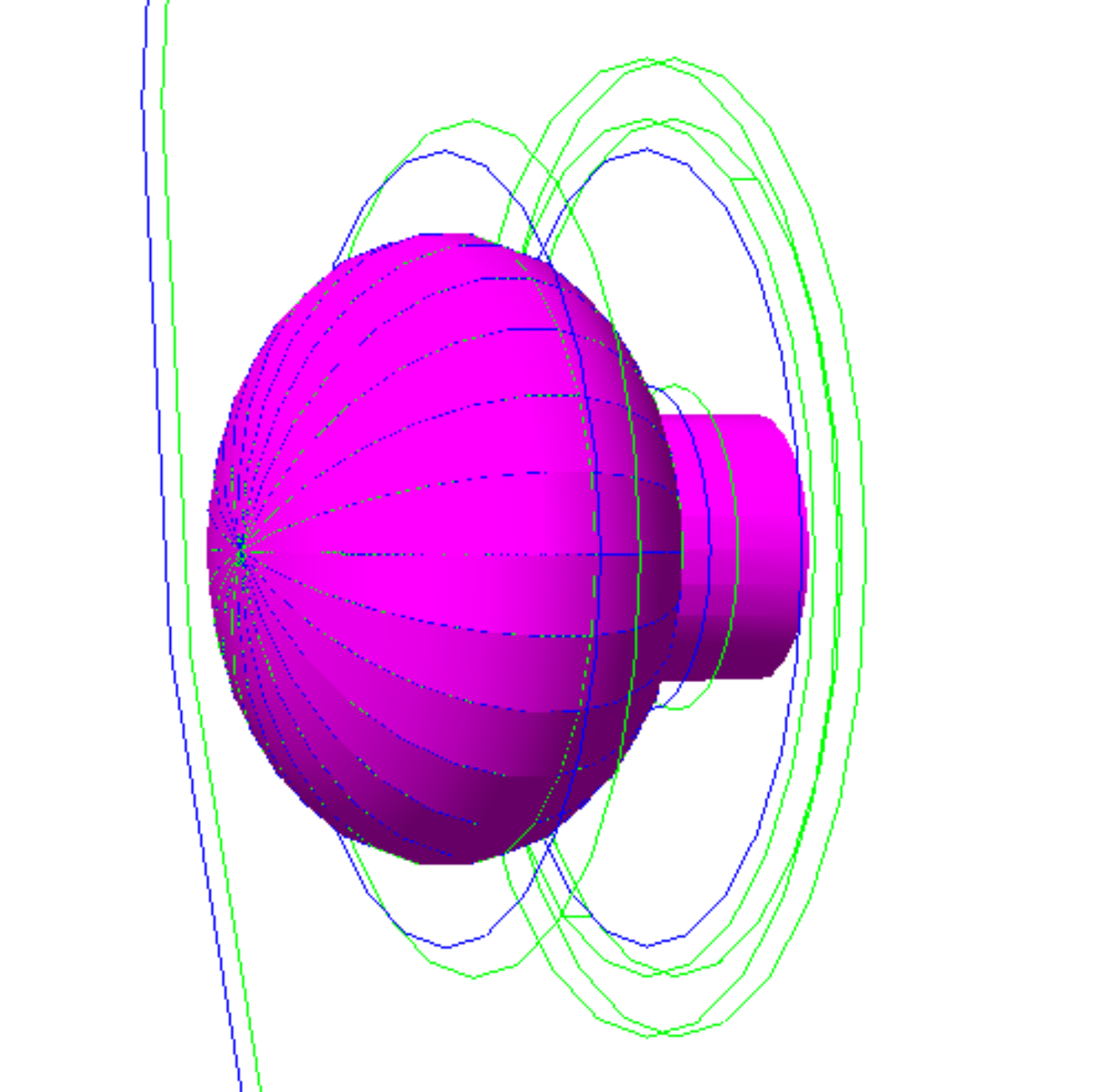}}
 \hspace{5mm}
 \subfigure[Steel vessel and Crystal Insertion System (CIS)]
   {\includegraphics[width=0.32\textwidth]{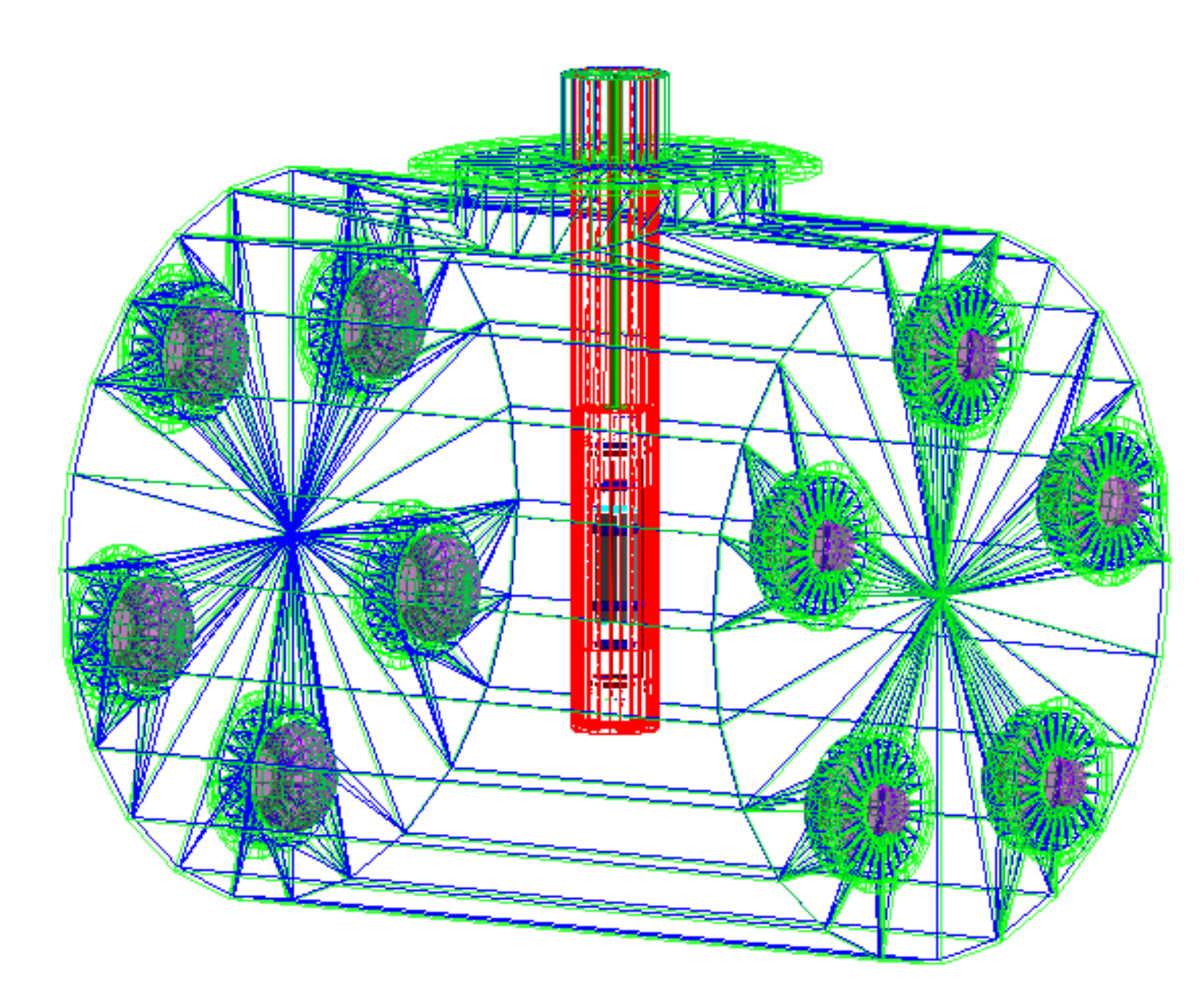}}
 \hspace{5mm}
 \subfigure[Full setup including external shielding]
   {\includegraphics[width=0.5\textwidth]{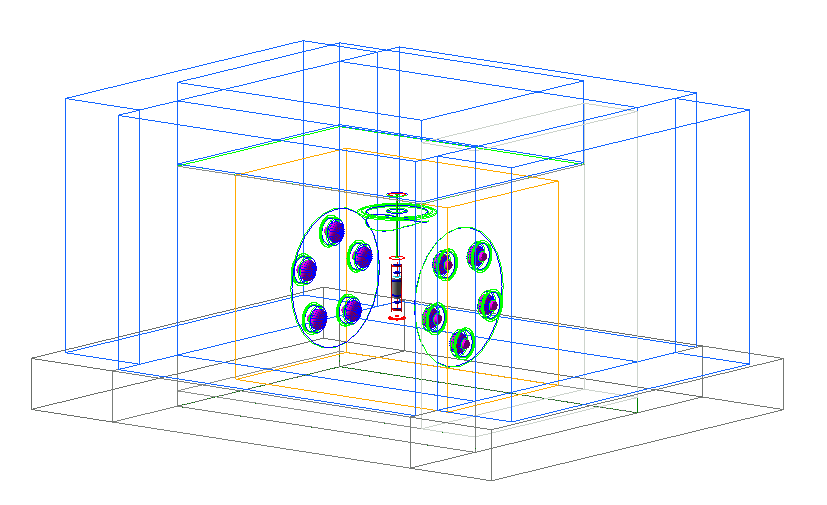}}
 \hspace{5mm}
 \caption{The SABRE--PoP setup as modeled and rendered in Geant4.}
 \label{fig:SimulatedGeometry}
 \end{figure}

The PoP setup consists of a crystal detector module, installed inside a cylindrical stainless steel vessel filled with liquid scintillator working as a veto against events with multiple interactions within the detector or from outside. The steel vessel is protected from the radioactivity in the experimental hall by a hybrid external shield made of lead, polyethylene and water.
The general design of the PoP experimental setup and of its sub-parts is described in~\cite{ConceptPaper}. Below we point out the main details and how they are implemented in the simulation, which uses the \textsc{Geant4} package~\cite{GEANT4,geant42006}, version 10.02 patch 3. 
%
The crystal detector module consists of a NaI(Tl) cylindrical crystal, with diameter 3.7 inches, length 8 inches, and mass 5.2 kg, coupled to two 3-inch diameter photomultiplier tubes (PMTs). The crystal is wrapped in a 200 $\mu$m thick Polytetrafluoroethylene (PTFE) foil. The 3" Hamamatsu R11065-20 PMTs are described by a simplified geometry, made of a quartz window, a Kovar body and a ceramic feedthrough plate, and directly coupled to the crystal. The whole assembly is placed inside a 2mm-thick cylindrical copper enclosure and held in position by PTFE holders and copper support rods. The external dimensions of the enclosure are 58 cm length and 14.6 cm diameter. A view of the internal parts of the enclosure, as implemented in simulation, is shown in Fig.~\ref{fig:SimulatedGeometry}(a). \\
The liquid scintillator veto consists of a cylindrical stainless steel vessel of 130 cm diameter $\times$ 150 cm length, filled with $\sim2$ tons of pseudocumene and viewed by ten 8" Hamamatsu R5912 PMTs. The PMTs are described by a simplified geometry: a photocathode ellipsoid plus a cylindrical body made of borosilicate glass, as shown in  Fig.~\ref{fig:SimulatedGeometry}(b). The mass of each PMT is about 1.1 kg. The crystal enclosure is inserted in a 0.2 cm thick cylindrical copper tube (16 cm diameter, 121 cm height) and connected to the top flange of the veto vessel through a stainless steel bar. The liquid scintillator (LS) veto and the Crystal Insertion System (CIS) are shown in Fig.~\ref{fig:SimulatedGeometry}(c). The external shielding is made of thick polyethylene (PE) slabs, arranged around the vessel on four sides. 
The compact polyethylene structure, whose thickness is 10 cm on the top and 40 cm on three sides, sustains the weight of a 2 cm steel plate and of water tanks placed on the top, for a total water thickness of 80 cm. The volume is closed by a polyethylene door 66 cm thick. All four polyethylene sides are surrounded by water tanks of 91 cm thickness. 
The bottom is further shielded by 10 cm of polyethylene and 15 cm of lead. The full setup, as implemented in the simulation, is shown in Fig.~\ref{fig:SimulatedGeometry}(d).

%

For the SABRE simulation, we have chosen the Shielding physics list recommended for underground low-background experiments, with the addition of the \textsc{Geant4} ``option 4" for the electromagnetic (EM) part ~\cite{geant42016}. 
The package for EM interactions includes the Wentzel VI model at high energy, Msc95 model below 100 MeV~\cite{wentzel}, photon models from Livermore and Penelope, and Livermore ionisation model for electrons \cite{Livermore, Penelope}. 
The hadronic interaction model includes elastic, inelastic, capture and fission processes; precision models are used for neutrons with energy below 20 MeV.
The production and transport of optical photons both in crystal and in the LS veto have not been included in the simulation results described here, however, their inclusion is being pursued currently.  
\section{Radioactive contamination of the SABRE--PoP materials}
\label{sec:contaminations}
The most relevant sources of radioactive contamination in the materials are primordial radionuclides ($^{238}$U, $^{232}$Th and their daughters and $^{40}$K), anthropogenic radionuclides (e.g.$\,^{137}$Cs), cosmogenic radionuclides and environmental radioactive noble gases, such as $^{222}$Rn and $^{220}$Rn.
The contamination levels of the materials composing the SABRE experiment are based on screening techniques such as gamma ray spectroscopy using High-Purity Ge (HPGe) detectors, neutron activation analysis (NAA), Accelerator Mass Spectroscopy (AMS), and Inductively Coupled Plasma Mass Spectroscopy (ICP-MS).\\
In sections \ref{sec:cont_crystal}-\ref{sec:cont_veto} we list the radioactive contamination of materials used as input to our Monte Carlo simulation. In some cases when the contamination level was below the sensitivity of the measurement, we consider the latter as an upper limit and conservatively use it in the simulation. Secular equilibrium in the U and Th decay chains is assumed, unless otherwise specified.\\
A summary of the experimental components implemented in the \textsc{Geant4} simulation with the corresponding materials and masses is reported in Table~\ref{tbl_masses}.

\begin{table}[htbp!]
\footnotesize
\centering
\begin{tabular}{|l|l|l|}
\hline
Volume Name & Material & Mass [kg]\\
\hline
 \multicolumn{3}{|l|}{{\bf Crystal}} \\
\hline
Crystal & NaI & 5.2 \\
 \hline
 \multicolumn{3}{|l|}{{\bf Enclosure}} \\
 \hline
Crystal Wrapping &  PTFE &  $2.6 \cdot 10^{-3}$\\
Enclosure body &  Copper &  $1.5 \cdot 10^{1}$ \\
Enclosure small parts &  Copper &   $1.2 $ \\
Enclosure small parts &  PTFE &   $3.2 \cdot 10^{-1}$ \\
 \hline
 \multicolumn{3}{|l|}{{\bf Crystal PMTs}} \\
 \hline
Window (x2) & Quartz  & 6.0 $\cdot 10^{-2}$ \\
Body (x2) & Kovar & 1.8 $\cdot 10^{-1}$\\
Feedthrough Plate (x2) &Ceramic & 3.1 $\cdot 10^{-2}$\\
 \hline
 \multicolumn{3}{|l|}{{\bf Crystal Insertion System (CIS)}} \\
 \hline
 CIS Tube & Copper & $1.2 \cdot 10^{1}$ \\
 CIS Bar & Stainless steel & $8.3 \cdot 10^{-1}$\\
 \hline
 \multicolumn{3}{|l|}{{\bf Veto}} \\
 \hline
 Scintillator & Pseudocumene & $1.9 \cdot 10^{3}$ \\
 Vessel & Stainless steel & $6.7 \cdot 10^{2}$ \\
 PMTs (x10)& Borosilicate glass & $1.1 \cdot 10^{1}$\\
\hline
 \multicolumn{3}{|l|}{{\bf Shielding}} \\
\hline
Walls and Top & PE & $11.5 \cdot 10^3$ \\
Base& PE & $7.7\cdot 10^3$ \\
Top & Stainless steel & $1.2\cdot 10^3$ \\
Base & Pb  & $15.6\cdot 10^3$ \\
Walls and Top & Water & $3.9\cdot 10^4$ \\ 
\hline
\end{tabular}
  \caption{Experimental components implemented in the \textsc{Geant4} code with corresponding materials and masses.}
  \label{tbl_masses}
\end{table}

\subsection{NaI(Tl) Crystals}
\label{sec:cont_crystal}
Contamination levels for the NaI(Tl) crystal in the simulation are taken either from measurements performed on the high-purity Astro Grade powder, or from the measurement of a $\sim$ 2-kg test crystal grown by RMD for the SABRE collaboration in 2015~\cite{ConceptPaper}. 
Uranium and Thorium contamination values in the powder were found below 1 ppt \cite{PNNLMeas} and showed no increase in the test crystal with respect to the starting powder.	
The average $^{39}$K level found in the crystal at different positions with two different methods was $9\pm1\,$ppb~\cite{PNNLMeas,ARNQUIST201715,SeastarMeas}. 
The concentration of $^{87}$Rb, which is another common long--living contaminant in NaI powder, is below the detection limit of 0.1 ppb in the grown crystal~\cite{SeastarMeas}.
Concerning the isotope $^{210}$Pb, this cannot be measured at the low levels required at present, although AMS methods are being developed. We rely on contamination levels measured by other NaI experiments, through the operation of their grown crystals as scintillators. The DAMA~\cite{DAMA2008} experiment finds that a background contribution from internal $^{210}$Pb is visible only in some of the crystals and quotes a range of activity levels of 0.005 - 0.03 mBq/kg. On the other hand, the ANAIS and COSINE experiments, who both have published their background models and compared with data~\cite{ANAIS2016,COSINEbkgmodel}, quote a much larger contamination, around 0.7 mBq/kg in their best crystals and up to 3.2 mBq/kg. 
In this work, we have used 0.03 mBq/kg, the highest level among the $^{210}$Pb contaminations found in DAMA crystals, as we are confident that the SABRE procedure for crystal growth and handling will minimize the $^{210}$Pb contamination. A quantitative discussion on the impact of an higher $^{210}$Pb contamination on the SABRE background is given in Section~\ref{sec:Results}, where we report the expected background per Bq/kg of $^{210}$Pb activity.
Cosmogenic--induced contaminations in the crystals are also relevant for a dark matter detector. Among these is $^3$H, especially worrisome as it decays beta releasing 18.6 keV of energy. The COSINE experiment determines a $^3$H activity in its NaI crystals at the level of 0.1 mBq/kg~\cite{COSINEbkgmodel}. Contaminations of tritium are difficult to predict prior to data taking. They indeed depend strongly on the history of the crystal and cannot be measured with HPGe. 
The cosmogenic activation in NaI crystals has been extensively studied by the ANAIS collaboration~\cite{ANAIS-COSMO}. 
We considered the list of cosmogenic activated isotopes produced in their NaI(Tl) crystals, and report their expected activities in Table~\ref{tblbulk_NaI}, together with the corresponding half lives. We used the ACTIVIA~\cite{ACTIVIA} simulation software to calculate the cosmogenic activation of these isotopes at sea level and during transport by plane, assuming an exposure at sea level of about 1 year plus a transport by plane from USA to Italy ($\sim 10$ hours of flight). We take into account a corrective factor for the geomagnetic effect. The resulting values, reported in Table~\ref{tblbulk_NaI}, have been used as input to our simulations. For $^{22}$Na only, we used the value measured with HPGe at LNGS on Astro Grade powder~\cite{laubenstein}, which is higher than the value simulated with ACTIVIA. For the long-lived $^{129}$I, which is treated as stable by ACTIVIA (15.7 million years half life), we have used the value measured by DAMA~\cite{DAMA2008} in their crystals.
When our own measurements of cosmogenic activation in the SABRE--PoP crystal operated as a scintillator will become available, we will use the simulated spectra with the correct activities to accurately model the cosmogenic background.


\begin{table}[htbp!]
\footnotesize
\centering
\begin{tabular}{|c|c|c|c|}
\hline
\multicolumn{4}{|c|}{Intrinsic} \\
\hline
	Isotope & \multicolumn{2}{|c|}{Activity [mBq/kg]} & Ref. \\
\hline
  $^{40}$K &   \multicolumn{2}{|c|}{0.31 }    & \cite{ConceptPaper} \\
  $^{238}$U &  \multicolumn{2}{|c|}{$<1.2 \cdot 10^{-2}$}     & \cite{ConceptPaper} \\
  $^{232}$Th & \multicolumn{2}{|c|}{$<4.1 \cdot 10^{-3}$}     & \cite{ConceptPaper} \\
  $^{87}$Rb &  \multicolumn{2}{|c|}{$<8.9 \cdot 10^{-2}$}   & \cite{ConceptPaper} \\
  $^{210}$Pb & \multicolumn{2}{|c|}{$<3.0 \cdot 10^{-2}$} & \cite{DAMA2008} \\
  $^{85}$Kr &  \multicolumn{2}{|c|}{$<1.0 \cdot 10^{-2}$} & \cite{DAMA2008} \\
\hline
\multicolumn{4}{|c|}{Cosmogenic} \\
\hline
  Isotope & Activity [mBq/kg] & Half life [days] & Ref. \\
\hline
  $^{3}$H     &    \centering 1.8 10$^{-2}$  &  \centering 4503   & \cite{ACTIVIA} \\ 
  $^{22}$Na     &    \centering 0.48  &  \centering 949   & \cite{laubenstein} \\ 
  $^{126}$I     &    \centering 4.1  &  \centering 13      & \cite{ACTIVIA} \\ 
  $^{129}$I     &    \centering 0.57  &  \centering -       &  \cite{DAMA2008}\\ 
  $^{113}$Sn    &    \centering 9.6 10$^{-2}$  &  \centering 115      &  \cite{ACTIVIA}\\ 
  $^{125}$I     &    \centering 1.9  &  \centering 59      &  \cite{ACTIVIA}\\ 
  $^{121m}$Te   &   \centering  0.50  &  \centering 154     & \cite{ACTIVIA} \\ 
  $^{123m}$Te   &    \centering 0.31  &  \centering 119     & \cite{ACTIVIA} \\ 
  $^{125m}$Te   &   \centering  0.69  &  \centering 57      & \cite{ACTIVIA} \\ 
  $^{127m}$Te   &    \centering 0.50  &  \centering 107     & \cite{ACTIVIA} \\ 
\hline
\end{tabular} 
\caption{Radioactivity levels assumed for the NaI(Tl) SABRE crystals. \label{tblbulk_NaI}}
\end{table}

\subsection{Crystal PMTs and reflector foil}
\label{sec:cont_PMT}
The radioactivity values of PMTs are based on measurements by the XENON Collaboration. They have performed extensive
HPGe screening of the Hamamatsu PMT R11410 which is identical to the model R11065 except for the photocathode material. Assuming no significant radioactivity can be attributed to the tiny amount of photocathode material, we have adopted the measurements (or upper limits) from~\cite{XenonPMTs} in our simulation.  The screening was performed on the main raw materials constituting the tubes and on batches of assembled PMTs. The results show that the radioactivity levels of assembled PMTs are not always compatible with the sum of the activities of the raw materials (expressed in [mBq/PMT]). For example, $^{40}$K and $^{60}$Co results were higher by about a factor 6 and 10 in the assembled PMT with respect to the sum of the parts. \\ 
In our simulations, we model the crystal PMTs as the assembly of three components. These are the three that give the highest contribution in terms of mass and radioactivity, namely the Kovar body, the quartz window and the ceramic feedthrough plates. The contamination values assigned to each of the three components are reported in Table~\ref{bulk_PMT} and have been calculated from the values in Table 3 and 4 of~\cite{XenonPMTs}, divided by the mass of the \textsc{Geant4} solid corresponding to the given PMT component\footnote{Masses of \textsc{Geant4} solids have been compared with Table 2 of the reference and found in agreement.}.  
To account for the higher radioactivity levels measured in the assembled PMT, the above values have been rescaled so that, for each isotope, the summed contribution from the three parts matches the total measured value from Table 5 of~\cite{XenonPMTs} and at the same time the ratios of activity levels in the three parts are kept constant and equal to those measured in the raw materials. \\
The contaminations of $^{235}$U and $^{137}$Cs are also reported in~\cite{XenonPMTs}, however, since they are upper limits or non-standard contaminations, they have not been considered in this work.

\begin{table}[h!]
\footnotesize
\centering
\begin{tabular}{|c|c|c|c|}
\hline
&  \multicolumn{3}{|c|}{Activity [mBq/PMT]} \\
\cline{2-4}
\multicolumn{1}{|c|}{Isotope}  & Body & Window & Ceramic plate \\
\hline

$^{40}$K      & $<$5.9	      &   $<$0.48	& 6.5	 \\ 
$^{60}$Co     & 0.65	      &   $<$0.042	& $<$0.19	 \\ 
$^{238}$U     & $<$0.52	      &   $<$1.8	& 13	 \\
$^{226}$Ra    & $<$0.29	      &   0.040	        & 0.29	 \\
$^{232}$Th    & $<$0.0098     &   $<$0.037      & 0.70	 \\ 
$^{228}$Th    & $<$0.41	      &   $<$0.015      & 0.13	 \\ 

\hline 
\end{tabular}
\caption{Radioactivity levels of PMT components, obtained by rescaling the values of raw materials from Table 3 and 4 of~\cite{XenonPMTs}, in order to obtain the same total PMT radioactivity as reported in their Table 5. \label{bulk_PMT}}
\end{table}

PMTs are coupled to the crystal by using Dow Corning optical silicone grease. Its contribution in the radioactivity background is ne\-gli\-gi\-ble. This statement is supported by the comparison of Dow Corning radioactivity level ~\cite{radiopurity} with the PMT window. Since both of materials are located close to the crystal they can be compared to each other. The PMT window has a radioactivity level 5 times higher than the optical grease for $^{238}$U, 3 times higher for  $^{232}$Th and 27 higher for $^{40}$K. Moreover, the mass of the PMT window is greater by about 60 times than the mass of optical grease that will likely be used for the coupling.

To evaluate the background contribution from the reflector material wrapped around the crystal, we assumed the contamination values listed in Table~\ref{bulk_PTFEWrapping}, measured by ICP-MS by the XENON experiment on a thin PTFE sheet used as a light reflector \cite{XENONScreening}.
\begin{table}[htbp!]
\footnotesize
\centering
\begin{tabular}{|c|c|}
\hline
Isotope & Activity [mBq/kg]  \\
\hline
$^{40}$K & 3.1   \\
$^{238}$U & 0.25 \\
$^{232}$Th & 0.5 \\
\hline
\end{tabular}
\caption{Radioactivity levels of PTFE reflector foil~\cite{XENONScreening}. \label{bulk_PTFEWrapping}}
\end{table}

\subsection{Copper and PTFE parts}
\label{sec:cont_enclosure}
The crystal enclosure and the crystal insertion system are mainly made of oxygen-free high-thermal-conductivity (OFHC) C10100 copper. The intrinsic background coming from copper parts has been evaluated taking into account the contamination of  $^{238}$U and $^{232}$Th decay chains and $^{40}$K. Since the copper used in the manufacturing of the enclosure comes from a batch also used by the CUORE experiment, activities were assumed to be at the same level of those measured by the CUORE collaboration \cite{CUORE2NU}. The values are reported in Table~\ref{cont_copper}. In the background simulation, we have also taken into account the cosmogenic radio-activation of copper. In a study performed for the XENON experiment the cosmogenic activation of OFHC copper was measured after 345 days of exposure to cosmic rays at 3470 m above sea level and the high-altitude activation measurements were then converted into specific saturation activities at sea level \cite{Baudis2015}. The values are reported in Table~\ref{cont_copper}.
 

\begin{table}[h!]
\footnotesize
\centering

\begin{tabular}{|c|c|c|c|}
\hline
\multicolumn{4}{|c|}{Intrinsic} \\
\hline
	Isotope & \multicolumn{2}{|c|}{Activity [mBq/kg]} & Ref. \\
\hline

$^{40}$K &   \multicolumn{2}{|c|}{0.7 } &   \cite{CUORE2NU}\\
$^{238}$U &   \multicolumn{2}{|c|}{0.065 } &  \cite{CUORE2NU} \\
$^{232}$Th &   \multicolumn{2}{|c|}{0.002 } &  \cite{CUORE2NU}\\
\hline
\multicolumn{4}{|c|}{Cosmogenic} \\
\hline
  Isotope & Activity [mBq/kg] & Half life [days] & Ref. \\
\hline
$^{60}$Co & 0.340 & 1925 & \cite{Baudis2015} \\
$^{58}$Co & 0.798 & 71 &  \cite{Baudis2015} \\
$^{57}$Co & 0.519 & 272 &  \cite{Baudis2015} \\
$^{56}$Co & 0.108 & 77 &  \cite{Baudis2015} \\ 
$^{54}$Mn & 0.154 & 312 &  \cite{Baudis2015} \\
$^{46}$Sc & 0.027 & 84 &  \cite{Baudis2015} \\
$^{59}$Fe & 0.047 & 44 &  \cite{Baudis2015} \\
$^{48}$V & 0.039 & 16 &  \cite{Baudis2015} \\
\hline
\end{tabular}
\caption{Radioactivity levels assumed for the copper parts of the SABRE crystal enclosure.\label{cont_copper}}
\end{table}

Several PTFE rings are used in the crystal enclosure to hold the crystal and the PMTs in place. The XENON collaboration reported on an extensive material screening campaign~\cite{XENONScreening}, where no evidence for radioactive contaminants in PTFE was present within the spectrometer sensitivity. We conservatively use
these upper limits (Table~\ref{bulk_PTFE}) as activity values in our simulations.

\begin{table}[h!]
\footnotesize
\centering
\begin{tabular}{|c|c|}
\hline
Isotope & Activity [mBq/kg] \\
\hline
$^{40}$K &  $<$2.25   \\
$^{238}$U & $<$0.31   \\
$^{232}$Th & $<$0.16 \\
$^{60}$Co & $<$0.11  \\
$^{137}$Cs & $<$0.13 \\
\hline
\end{tabular}
\caption{Radioactivity levels assumed for the PTFE parts of the SABRE crystal enclosure. Values are taken from \cite{XENONScreening}. \label{bulk_PTFE}}
\end{table}

\subsection{Veto components: Stainless steel, PMTs and Liquid Scintillator}
\label{sec:cont_veto}
Stainless steel is used in the veto vessel and in the crystal insertion system. The SABRE--PoP veto vessel was manufactured by Allegheny Bradford Corporation (ABC) in Bradford, PA, USA. The main components are a 1/4'' thick plate that was rolled into the cylinder and a 3/8'' thick plate that was used for the side walls, the top flange and the top plate.
Table~\ref{bulk_steel} summarizes the radioactivity of steel samples from Stainless Plate Products (SPPUSA) measured using the glow discharge mass spectrometry (GDMS) method. The most conservative values (higher contamination) have been used for the steel vessel and the CIS steel bar.
\begin{table}[htbp!] 
\footnotesize
\centering
\begin{tabular}{|c|c|c|}
\hline
&  \multicolumn{2}{|c|}{Activity [mBq/kg]} \\
\cline{2-3}
\multicolumn{1}{|c|}{Isotope}  & Lot n.S536 & Lot n.T915\\
 & Thickness 3/8" & Thickness 1/4" \\
\hline
$^{40}$K & 0.12  & $<$ 0.03\\
$^{238}$U &  3.7  & 0.49\\
$^{232}$Th & $<$0.41 & 0.082\\
\hline 
\end{tabular}
\caption{Radioactivity inferred from concentration measurements with GDMS of stainless steel used for the SABRE--PoP vessel and CIS~\cite{ShieldsPhD}.  \label{bulk_steel}}
\end{table}

The active veto uses ten 8'' Hamamatsu R5912 PMTs. The PMT is made from low--radioactivity borosilicate glass and the corresponding contaminations (Table~\ref{bulk_vetopmt}) have been measured by the DarkSide--50 collaboration, which uses the same model~\cite{DS50PMT}. 

\begin{table}[htbp!]
\footnotesize
\centering
\begin{tabular}{|c|c|}
\hline
Isotope & Activity [mBq/PMT] \\
\hline
$^{40}$K & 649\\ 
$^{238}$U & 883\\
$^{232}$Th & 110\\
$^{235}$U & 41 \\
\hline
\end{tabular}
\caption{Radioactivity levels of the Hamamatsu R5912 Veto PMTs. Values are taken from \cite{DS50PMT}. \label{bulk_vetopmt}}
\end{table}

The SABRE--PoP liquid scintillator veto consists of $\sim$2 ton of high-purity pseudocumene (PC) doped with 3 g/l PPO as wavelength shifter. The scintillator will be provided by the Borexino experimental facilities. Consequently we have adopted the contamination levels
measured by Borexino~ \cite{BorexinoLS}, as reported in Table~\ref{bulk_ls}.
\begin{table}[htb]
\footnotesize
\centering
\begin{tabular}{|c|c|c|}
\hline
Isotope & Activity [mBq/kg] \\
\hline
$^{40}$K &  $3.5 \cdot 10^{-7}$  \\
$^{238}$U & $< 1.2\cdot 10^{-6}$ \\
$^{232}$Th & $< 1.2\cdot 10^{-6}$  \\
$^{210}$Pb & $1.7\cdot 10^{-6}$   \\
$^{210}$Bi & $1.7\cdot 10^{-6}$   \\
$^{7}$Be & $< 1.2\cdot 10^{-6}$   \\
$^{14}$C & $4.1\cdot 10^{-1}$   \\
$^{39}$Ar & $3.5\cdot 10^{-6}$   \\
$^{85}$Kr & $3.5\cdot 10^{-7}$   \\
\hline
\end{tabular}
\caption{Radioactivity levels assumed for the liquid scintillator. Values are taken from \cite{BorexinoLS}. \label{bulk_ls}}
\end{table}
\subsection{External Shield}
\label{sect:ext_shield}
The shielding geometry is implemented in the simulation but is treated only as a passive material, i.e. radioactivity of the shielding and its contribution to the background budget of the PoP is not included in this work. Previous simulations performed with the same Monte Carlo code to support the shielding design show that a radio-purity at the level measured at LNGS, by HPGe spectroscopy on samples of SABRE polyethylene (K and gamma emitters from U/Th chain), gives a negligible contribution to the total background.\\
The same is true for the contribution from radiogenic neutrons in the LNGS rock, in the passive shielding materials and in the liquid scintillator. A preliminary estimation of this contribution was performed using the SOURCES~\cite{SOURCES} code, the results of which were used as an input to the Geant4 si\-mu\-la\-tion.

\section{Results}\label{sec:Results}
We simulated radioactive decays in different components of the setup, according to the radioactive contaminations of the materials described in the previous sections.
For every combination of isotope and location inside the setup, we simulated a number of events high enough to keep the statistical uncertainty in the output spectrum well below the percent level for the crystal background and below a few percent for the outermost volumes.  

The crystal and the liquid scintillator are treated in the \textsc{Geant4} simulation as sensitive detectors, and all the energy deposited by radiated particles inside those volumes is recorded.
The optical simulation, meaning the generation, propagation and collection of optical photons from scintillation, is not carried out in this work. 
We applied a gaussian smearing to the data to account for a $2 \% \cdot\sqrt{E}$ [MeV] resolution on the reconstructed energy in the NaI(Tl) crystal, while an energy resolution of $6 \% \cdot\sqrt{E}$ [MeV] has been considered for the veto signal. These are representative values for detectors such as NaI(Tl) crystals and Pseudocumene+PPO liquid scintillator detectors \cite{DAMA2008,BorexinoLS}.
The detection efficiency of the liquid scintillator veto is assumed to be 100\% for energy deposited above 100 keV \cite{ShieldsPhD}. \\
We evaluated the background contributions in the two distinct operation modes that we anticipate using during the PoP operation: the potassium Measurement Mode (KMM) and the Dark Matter Measurement Mode (DMM), where the liquid scintillator detector is used in coincidence or anti-coincidence with the crystal, respectively.


\subsection{Expected background in potassium measurement mode (\kmm)}\label{sec:KMM}
\begin{figure}[ht!] \centering
  \includegraphics[width=0.42\textwidth]{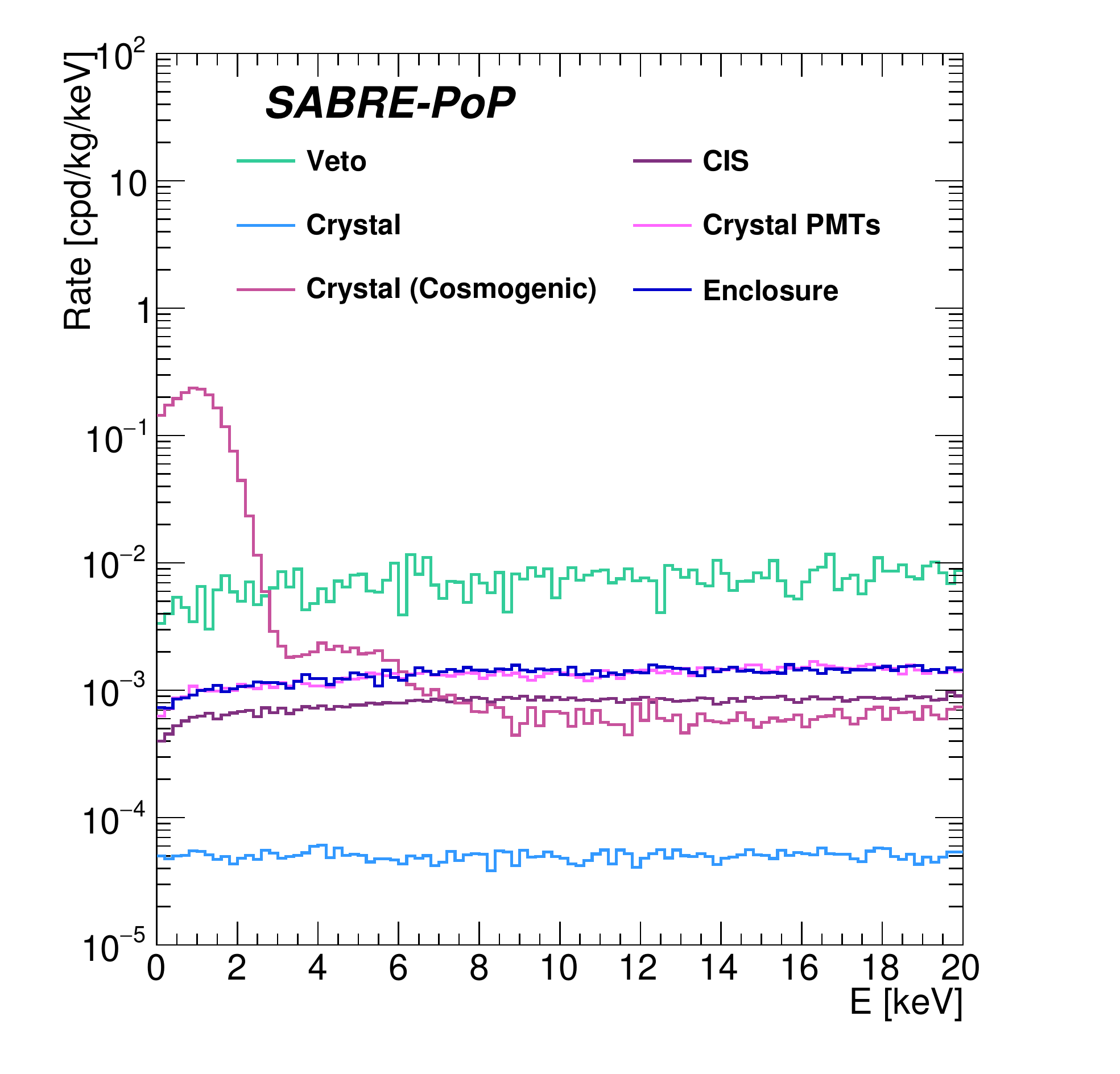}
  \includegraphics[width=0.42\textwidth]{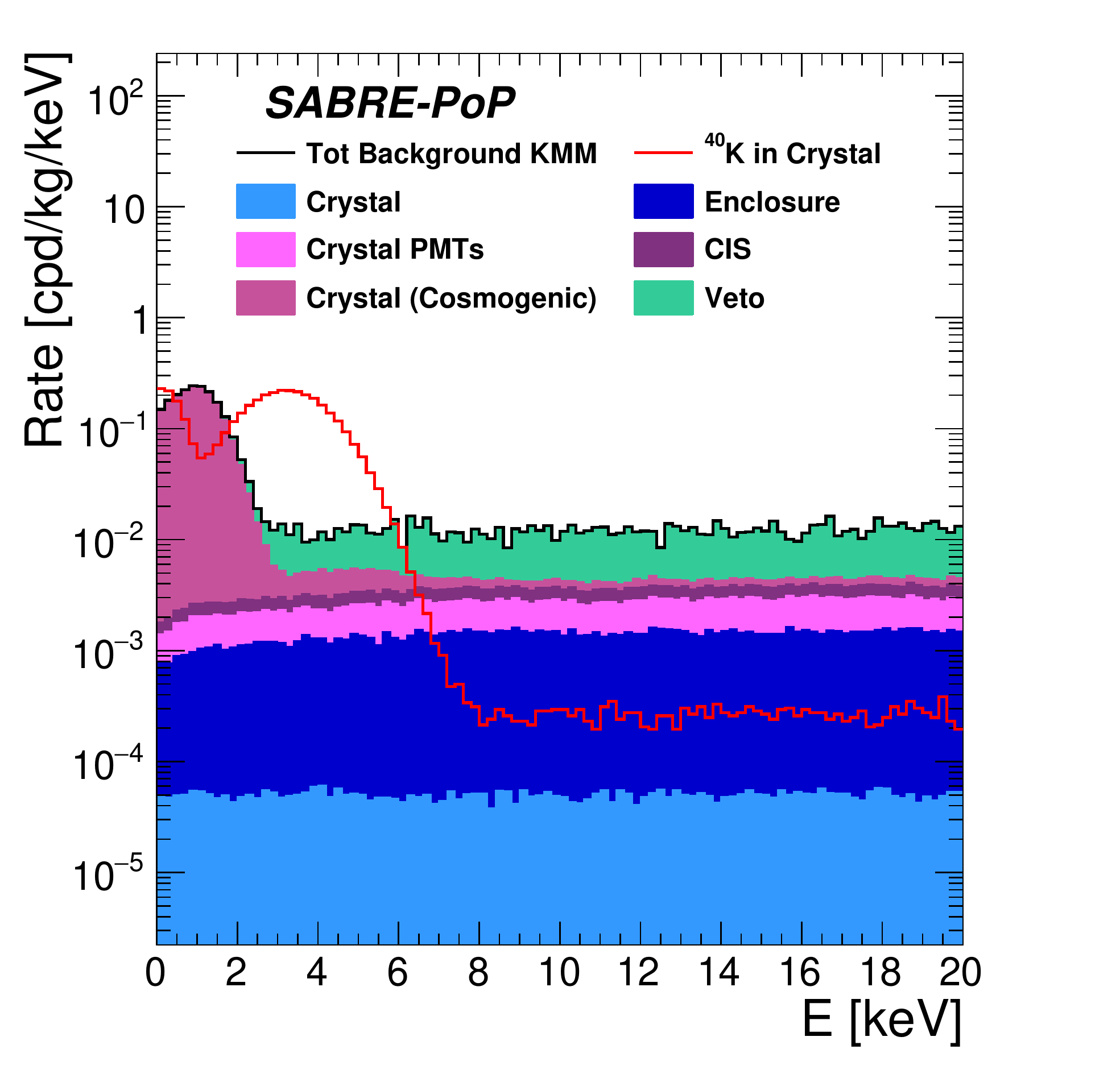}\\
  \caption{Backgrounds from all SABRE--PoP setup components in a $[0-20]$~keV region in potassium measurement mode (\kmm). The top plot shows the individual contributions from crystal (cosmogenics after 60 days underground and intrinsic backgrounds except $^{40}$K ), phototubes (PMTs), crystal enclosure, insertion system (CIS), and veto (including liquid scintillator, steel vessel and PMTs). The bottom plot shows the stacked backgrounds, and the total (solid black line). The red superimposed line shows the $^{40}$K spectrum from a 10 ppb $^{nat}K$ contamination in the crystal, which represents the signal in this configuration.}
  \label{fig:KMM60_lowEne}
\end{figure}

The \kmm will be used in the PoP to measure the $^{40}$K activity in the crystal, which is a significant background contribution to the energy spectrum in the region of interest for a dark matter search.
The electron capture (EC) from potassium decay in the crystal (about 11\% branching ratio) gives an energy deposit around 3.2 keV due to X-ray or Auger de-excitation, in coincidence with a $\gamma$ of 1.46 MeV. We assess the sensitivity of the SABRE--PoP to the measurement of the $^{40}$K activity in the crystal by evaluating the signal-to-background ratio in simulated data. The $^{40}$K signal is defined as an energy deposit between 2 and 4 keV in the crystal (1$\sigma$ around the 3 keV $^{40}$K peak) in coincidence with an energy deposit between 1.28 MeV and 1.64 MeV in the liquid scintillator (2.5$\sigma$ around the 1.46 MeV $^{40}$K peak). True coincidences from radioactive sources other than $^{40}$K contaminations in the crystal can mimic the same signature and therefore have to be regarded as background for the potassium measurement.\\
The background contributions from all the SABRE--PoP components to the crystal energy spectrum in \kmm are reported 
in Fig.~\ref{fig:KMM60_lowEne}. Cosmogenic activation is evaluated after 60 days underground. 
The signal produced by a 10 ppb $^{nat}K$ contamination in the crystal is superimposed in red in Fig.~\ref{fig:KMM60_lowEne}(b) for comparison. 

\noindent The crystal cosmogenic activation gives the most relevant contribution to the background in this measurement mode. 
This is mostly due to $^{22}$Na contamination, since $^{22}$Na emits 1275 keV $\gamma$ rays that often leave a deposit in the energy region selected for potassium.
The second-highest contribution to the background in KMM comes from the veto (namely the sum of scintillator, steel vessel and veto PMTs).

As shown in Table~\ref{bulk_kmm_total}, the total background is about one order of magnitude lower than the signal given by a 10~ppb $^{nat}K$ contamination in the crystal, thus demonstrating that the measurement of such potassium level is possible with $\sim$1~ppb precision in about two months of data taking. 

\begin{table}[htbp!]
\footnotesize
\centering
\begin{tabular}{|l|c|}
\hline
        &  Rate \kmm    \\
        &  [cpd/kg/keV]\\
\hline
  Crystal Cosmogenic         &     $9.8 \cdot 10^{-3}$\\
  Veto                          &     $6.2 \cdot 10^{-3}$\\  
  Enclosure                  &     $1.3 \cdot 10^{-3}$\\
  Crystal PMTs                   &     $1.1 \cdot 10^{-3}$\\
  CIS                        &     $7.7 \cdot 10^{-4}$\\
  Crystal (no $^{40}$K)         &     $5.1 \cdot 10^{-5}$\\
\hline
  Total                         &     $2.5 \cdot 10^{-2}$\\
  \hline
  \textbf{Crystal $^{40}$K}     &     \textbf{$1.9 \cdot 10^{-1}$}\\
\hline
\end{tabular}
  \caption{Background rates in the region $[2-4]$~keV in \kmm from all the SABRE--PoP setup components, listed in decreasing order. The signal of $^{40}$K in the crystal is reported below the total background and in bold. Cosmogenic backgrounds for crystal, enclosure and CIS are computed after 60 days underground.}
  \label{bulk_kmm_total}
\end{table}

\subsection{Expected background in dark matter measurement mode (\dmm)}\label{sec:DMM}
\begin{figure}[ht!] \centering
  \includegraphics[width=0.42\textwidth]{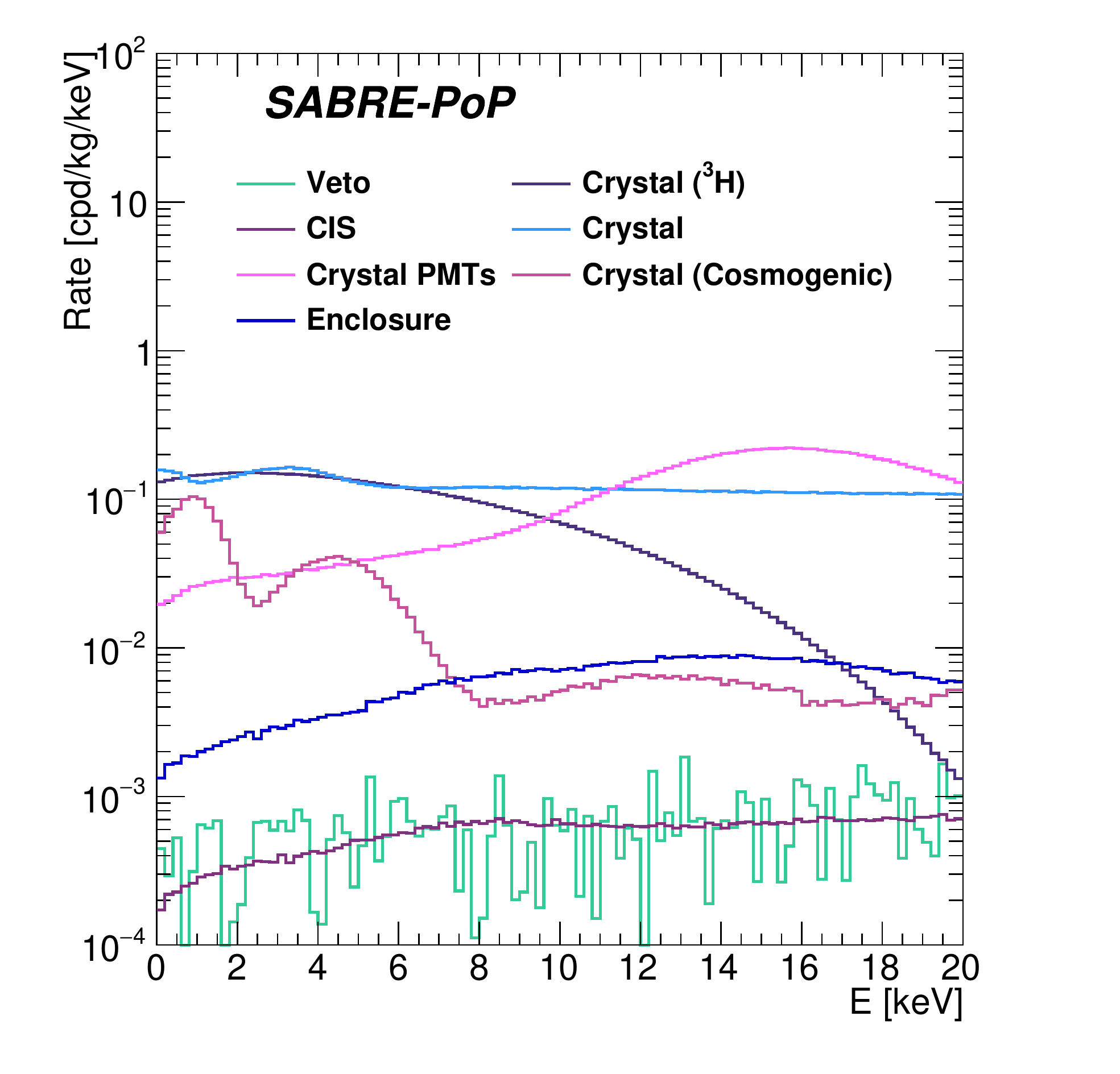}
  \includegraphics[width=0.42\textwidth]{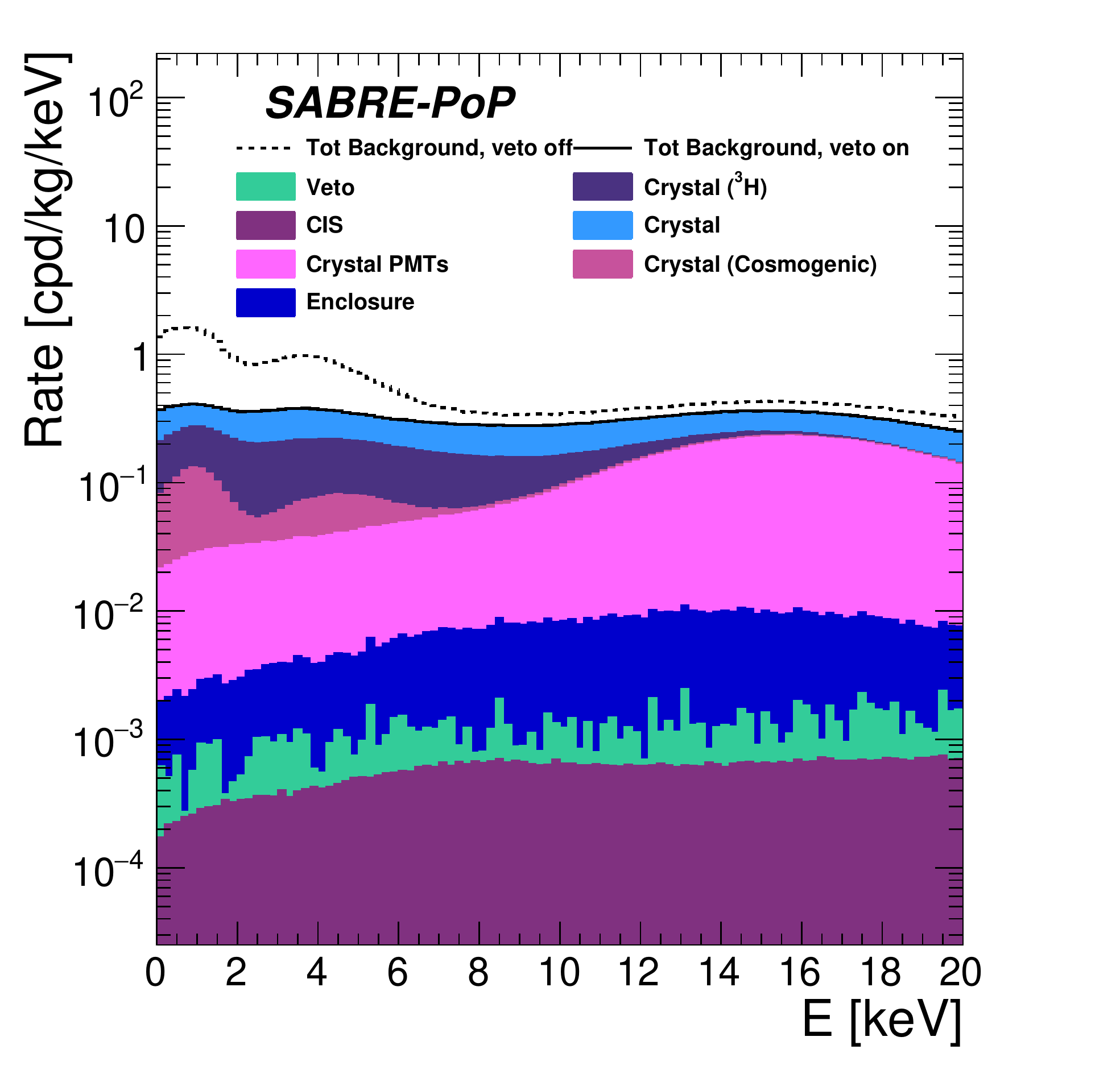}
  \caption{Backgrounds from all SABRE--PoP setup components in a $[0-20]$~keV region in dark matter measurement mode (\dmm). The top plot shows, with veto on, the separate contributions from crystal (intrinsic backgrounds, $^{3}$H, and cosmogenics after 180 days underground), phototubes (PMTs), crystal enclosure, insertion system (CIS), and veto (including liquid scintillator, steel vessel and PMTs). The bottom plot shows the stacked backgrounds and the total with veto on (solid black line) and veto off (dashed black line).}
 \label{fig:DMM180_lowEne}
\end{figure}

\begin{figure}[htbp!] \centering
  \includegraphics[width=0.42\textwidth]{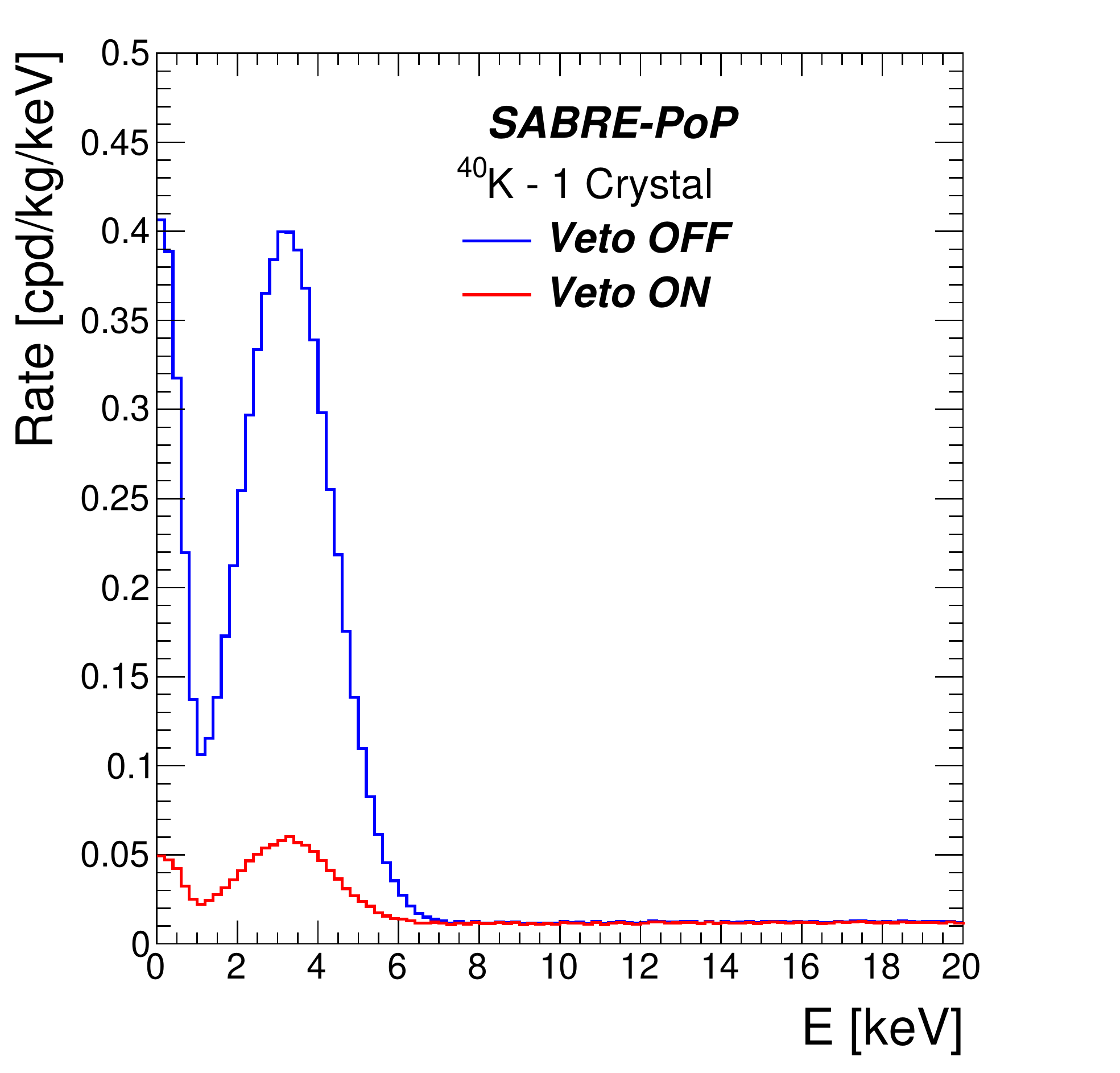}
  \caption{Effect of the LS veto on the $^{40}$K background in the crystal.}
  \label{fig:KVETO}
\end{figure}

We estimate the background level attainable by the SABRE--PoP in its planned configuration by looking at energy depositions from 2 to 6 keV in the crystal, in anti-coincidence with the veto. Dark matter interactions are single-site events that give rise to only one energy release. This is not the case for some of the background components, for example the intrinsic $^{40}$K described above. If the high-energy 1.46 MeV gamma from potassium decay escapes undetected from the crystal volume, the remaining 3.2 keV deposit in the crystal contaminates the low-energy region where a dark matter signal is expected. This can be avoided if the 1.46 MeV $\gamma$ is detected in the liquid scintillator veto. \\
The background contributions from all of the SABRE--PoP components to the crystal energy spectrum in \dmm is reported in Fig.~\ref{fig:DMM180_lowEne} and the background rates in the region of interest $[2-6]$~keV are reported in Table~\ref{bulk_dmm_total}, highlighting the effect of the liquid scintillator veto. 
\noindent The contribution from cosmogenic activation is evaluated assuming the data taking will start 180 days after the crystal is brought underground. The background from $^{3}$H is reported independently as it's the strongest source among cosmogenic isotopes and cannot be vetoed. Apart from this, the most relevant contribution in the 2--6 keV energy region comes from $^{121}$Te. Despite its short half life ($T_{1/2}$~=~17~days), the isotope is regenerated by the presence of the metastable $^{121m}$Te ($T_{1/2}$~=~154~days) which decays to the ground state with an internal transition probability of 0.886. \\
The rejection efficiency of the veto can be evaluated as the ratio of the number of events that give an energy deposition in the crystal in coincidence with a release of at least 100 keV in the liquid scintillator, over the total number of events in the crystal. 
For the $^{40}$K background in the crystal, in the region of interest for DM search [2-6] keV, the expected veto efficiency is 84\% (see also Fig.~\ref{fig:KVETO}). 
Overall the expected background rejection due to the liquid scintillator veto is 44\%, which is largely affected by the non-vetoed contribution from $^{3}$H. 

\begin{table}[htbp!]
\footnotesize
\centering
\begin{tabular}{|l|c|c|}
\hline
        & Rate, veto OFF & Rate, veto ON\\
        & [cpd/kg/keV]   & [cpd/kg/keV]\\
\hline

  Crystal            &   $3.5 \cdot 10^{-1}$ &    $1.5 \cdot 10^{-1}$\\
  Crystal ($^{3}$H)     &   $1.4 \cdot 10^{-1}$ &    $1.4 \cdot 10^{-1}$\\   
  Crystal Cosmogenic     &   $2.4 \cdot 10^{-1}$ &    $3.1 \cdot 10^{-2}$\\   
  Crystal PMTs           &   $4.3 \cdot 10^{-2}$ &    $3.5 \cdot 10^{-2}$\\
  Enclosure          &   $9.5 \cdot 10^{-3}$ &    $3.6 \cdot 10^{-3}$\\
  Veto                  &   $3.0 \cdot 10^{-2}$ &   $5.7 \cdot 10^{-4}$\\  
  CIS                &   $3.7 \cdot 10^{-3}$ &   $4.6 \cdot 10^{-4}$\\

\hline
  Total                 &   $8.2 \cdot 10^{-1}$ &   $3.6 \cdot 10^{-1}$\\
\hline
\end{tabular}
  \caption{Background rate in the region of interest $[2-6]$~keV from all the SABRE--PoP setup components with veto off and on respectively. The contributions are listed in decreasing order with veto on. Cosmogenic backgrounds for crystal, enclosure and CIS are computed after 180 days underground.}
  \label{bulk_dmm_total}
\end{table}

\begin{figure}[t!] \centering
  \includegraphics[width=0.42\textwidth]{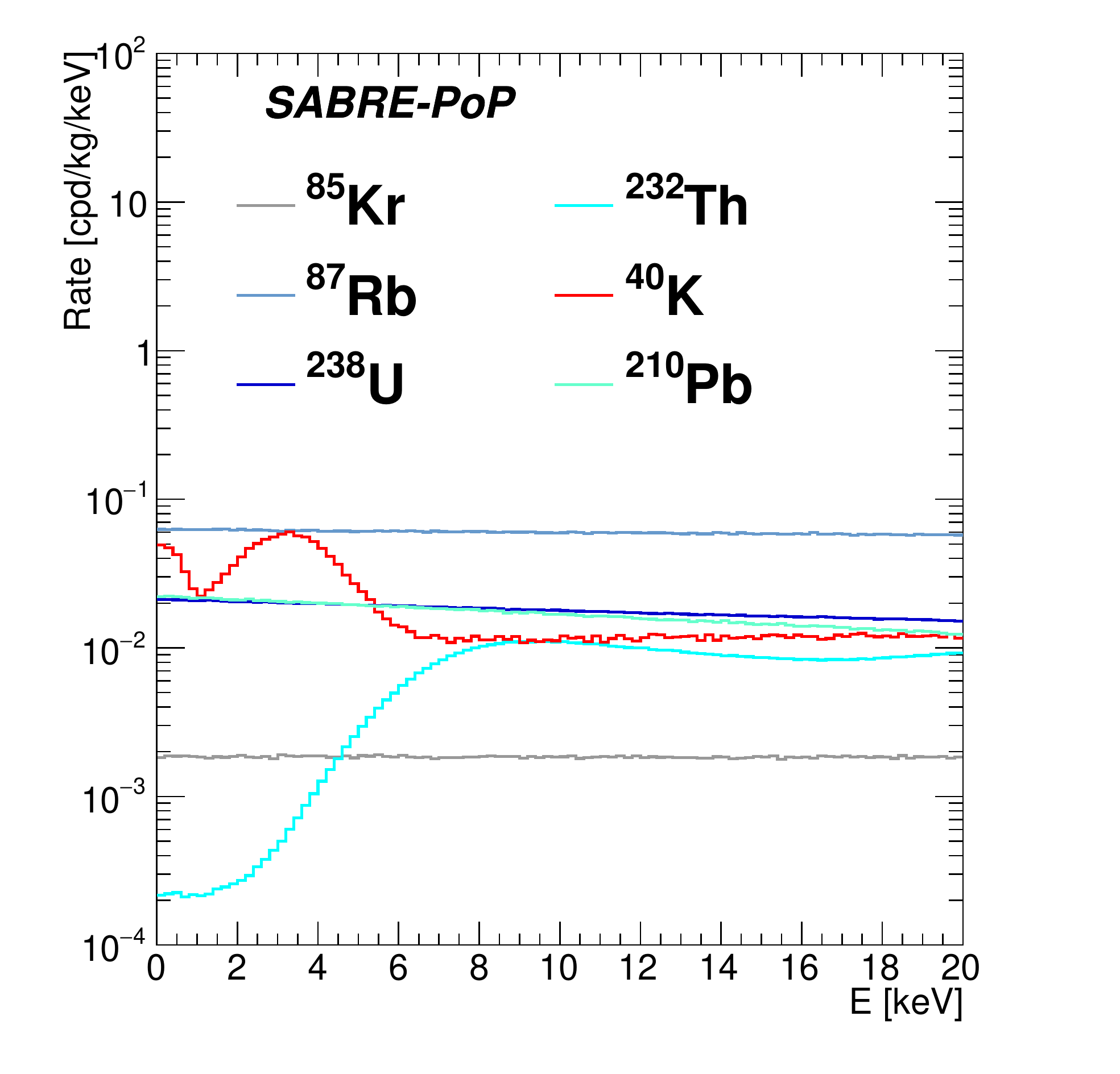}
  \includegraphics[width=0.42\textwidth]{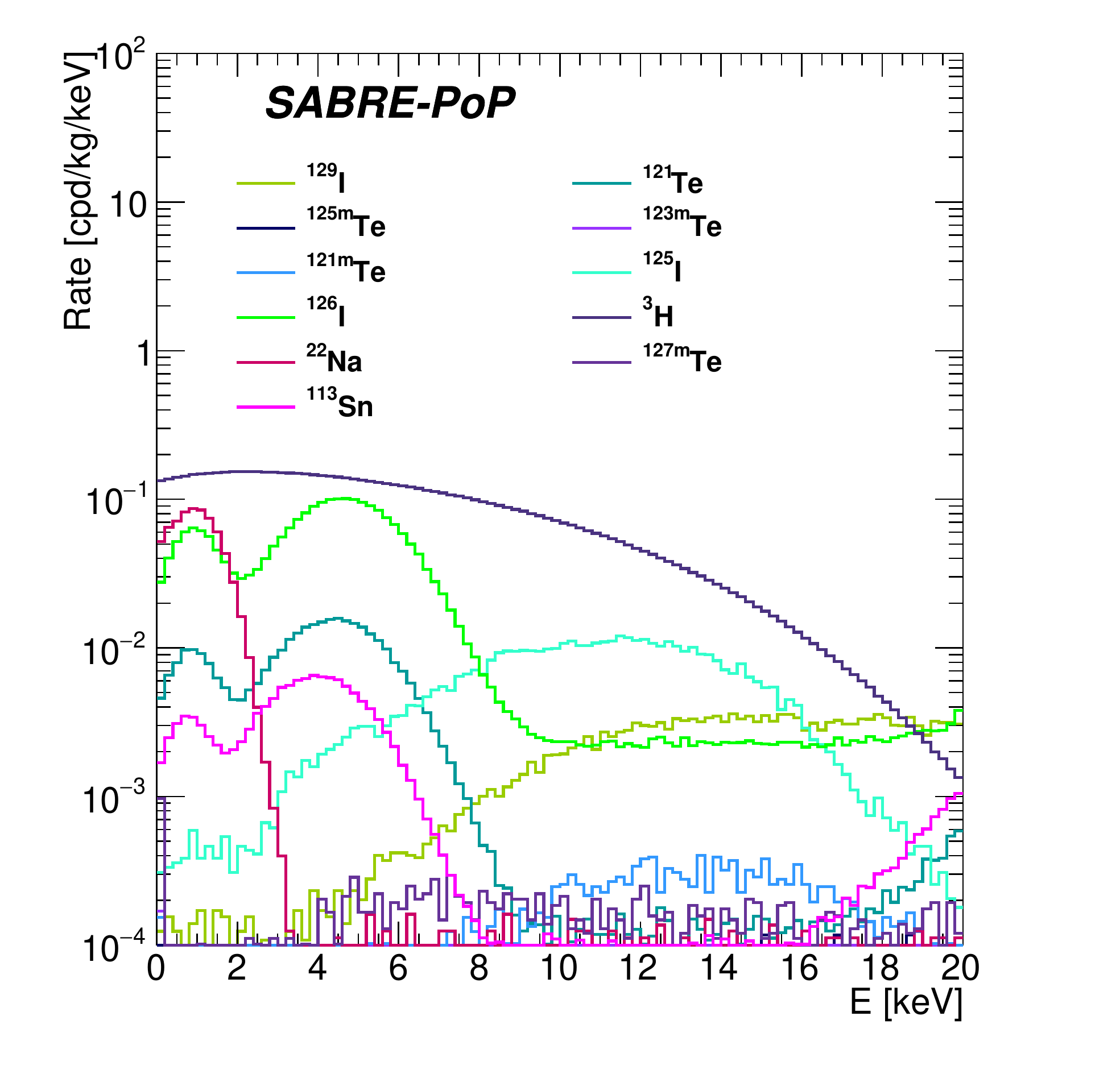}\\
  \caption{Background in a $[0-20]$~keV region due to the contaminants in the NaI(Tl) crystal. Top plot shows the contributions of intrinsic backgrounds. Bottom plot shows the contributions from cosmogenically-activated nuclei after 180 days underground. Veto is applied to the events that release more than 100~keV in the liquid scintillator.}
  \label{fig:crystal_lowEne}
\end{figure}

\noindent The simulated radioactive plus cosmogenic activated contamination of the crystal gives the most significant contribution in dark matter measurement mode. This confirms what was already estimated in previous SABRE simulation studies~\cite{ShieldsPhD}.
In Fig.~\ref{fig:crystal_lowEne} the background in a $[0-20]$~keV region due to the intrinsic and cosmogenic contaminations in the NaI(Tl) crystal is reported. The background rates in the region of interest $[2-6]$~keV are reported in Table~\ref{bulk_NaI}. 

\noindent Among the intrinsic contaminations listed in Table~\ref{bulk_NaI}, the most relevant source of background is $^{87}$Rb, followed by $^{40}$K. The $^{87}$Rb contamination, however, is assumed equal to the upper limit from the ICPMS measurement. 
Cosmogenic activation, mainly from $^{3}$H, has a significant impact on the total background. The total background contribution of this component ($1.7 \cdot 10^{-1}$ cpd/kg/keV with veto on) relies on calculations that assume an exposure at sea level of about 1 year plus a transport by plane from USA to Italy, followed by 180 days underground. The ultimate level of cosmogenic activation in the SABRE--PoP crystal will depend on how long the powder and the crystal are exposed to the cosmic ray flux. The results of our simulations highlight that minimising this exposure is crucial.
As discussed in Section~\ref{sec:contaminations}, the isotopes $^3$H and $^{210}$Pb contribute significantly to the count rate in the low energy region. Their activity in the SABRE--PoP crystal is difficult to predict at this stage and measuring them is among the goals of the PoP run. From simulations, each $\mu$Bq/kg of $^3$H activity in the crystal leads to $8.1 \cdot 10^{-3}$~cpd/kg/keV of background. At the same time, we estimate that, for the SABRE--PoP detector, 1~$\mu$Bq/kg of $^{210}$Pb will give a rate in the signal region of 0.67$\cdot 10^{-3}$~cpd/kg/keV. In the budget of Table~\ref{bulk_NaI} we have allowed for a $^3$H activity of 0.018 mBq/kg and a $^{210}$Pb activity of 0.03 mBq/kg. 

\begin{table}[ht!]
\footnotesize
\centering
\begin{tabular}{|c|c|c|}
\hline
Isotope & Rate, veto OFF & Rate, veto ON\\
        & [cpd/kg/keV]   & [cpd/kg/keV]\\
\hline
\multicolumn{3}{|c|}{Intrinsic} \\
\hline
  $^{87}$Rb  &  $6.1 \cdot 10^{-2}$ & $6.1 \cdot 10^{-2} $ \\
  $^{40}$K   &  $2.5 \cdot 10^{-1}$ & $4.0\cdot 10^{-2}$ \\
  $^{238}$U  &  $2.0 \cdot 10^{-2}$ & $2.0 \cdot 10^{-2}$ \\
  $^{210}$Pb &  $2.0 \cdot 10^{-2}$ & $2.0 \cdot 10^{-2} $ \\
  $^{85}$Kr &   $1.9 \cdot 10^{-3}$ & $1.9 \cdot 10^{-3} $ \\
  $^{232}$Th &  $1.9 \cdot 10^{-3}$ & $1.7 \cdot 10^{-3} $ \\

\hline
  Tot Intrinsic & $3.5 \cdot 10^{-1}$ & $1.4 \cdot 10^{-1}$  \\
\hline
\multicolumn{3}{|c|}{Cosmogenic}\\
\hline
  $^{3}$H        &    $1.4 \cdot 10^{-1}$   &  $1.4 \cdot 10^{-1}$  \\ 
  $^{121}$Te     &    $2.0 \cdot 10^{-1}$   &  $2.6 \cdot 10^{-2}$  \\ 
  $^{113}$Sn     &    $1.2 \cdot 10^{-2}$   &  $2.2 \cdot 10^{-3}$  \\ 
  $^{22}$Na      &    $2.1 \cdot 10^{-2}$   &  $1.5 \cdot 10^{-3}$  \\ 
  $^{125}$I      &    $4.4 \cdot 10^{-4}$   &  $4.4 \cdot 10^{-4}$  \\ 
  $^{129}$I      &    $1.9 \cdot 10^{-4}$   &  $1.9 \cdot 10^{-4}$  \\ 
  $^{126}$I      &    $1.8 \cdot 10^{-4}$   &  $1.2 \cdot 10^{-4}$  \\ 
  $^{127m}$Te    &    $6.4 \cdot 10^{-5}$   &  $6.4 \cdot 10^{-5}$  \\ 
  $^{121m}$Te    &    $7.1 \cdot 10^{-5}$   &  $3.7 \cdot 10^{-5}$  \\ 
  $^{123m}$Te    &    $1.9 \cdot 10^{-5}$   &  $1.3 \cdot 10^{-5}$  \\ 
  $^{125m}$Te    &    $3.8 \cdot 10^{-6}$   &  $3.7 \cdot 10^{-6}$  \\ 
\hline                     
  Tot Cosmogenic    &   $3.8 \cdot 10^{-1}$    &  $1.7 \cdot 10^{-1}$ \\ 
(180 days)    &     &   \\ 
\hline
\end{tabular}
\caption{Background rate in the region of interest $[2-6]$~keV due to the contaminants in NaI(Tl) crystals. Both intrinsic and cosmogenically-activated contributions are reported, with veto off and on respectively. The contributions are listed in decreasing order with veto on. The $^{87}$Rb contamination is an upper limit from ICPMS measurement. 
Cosmogenic backgrounds are computed after 180 days underground.}
\label{bulk_NaI}
\end{table}


\noindent As stated in Sect.~\ref{sect:ext_shield} the background budget due to the shielding materials has not been included in this work, since previous simulations performed with the same Monte Carlo have shown that the contamination of samples of SABRE polyethylene produce a background well below 10$^{-3}\,$cpd/kg/keV in dark matter measurement mode and therefore can be neglected. A preliminary simulation of radiogenic neutrons in the LNGS rock, in the passive shielding materials and in the liquid scintillator has shown that this contribution is at the level of 10$^{-4}\,$cpd/kg/keV in dark matter measurement mode.

\section{Conclusions}
\label{sec:conclusion}
The SABRE--PoP operation is expected to start at LNGS in 2018, with the goal of demonstrating that the intrinsic NaI(Tl) crystal contamination levels and the other backgrounds from the SABRE apparatus itself are low enough to carry out a reliable test of the DAMA result in the projected SABRE full-scale experiment. 
\noindent We evaluated the expected background of the SABRE--PoP with a Monte Carlo simulation based on the current knowledge of the most relevant sources of radioactive contamination. 
The simulation carefully reproduces the design of the PoP apparatus, with particular attention to the parts close to the crystal detector.  
We find that the radioactive contamination of the crystal gives the most significant contribution to the low-energy background, confirming the importance in the SABRE strategy of lowering the crystal contamination as much as possible. The radio-purity of the crystals, combined with the active veto technique, allows SABRE to achieve a design background of 0.36 cpd/kg/keV in the 2--6 keV energy region, where the maximum amplitude of the modulation was observed by DAMA. It is important to note that this estimation was derived under the assumptions that the intrinsic activities of $^3$H and $^{210}$Pb are below 0.018 mBq/kg and 0.03 mBq/kg respectively. Under these assumptions, the highest contribution to the crystal intrinsic background is $^{87}$Rb, for which we assumed the upper limit contamination. SABRE aims to be the first NaI(Tl)--based experiment with a background significantly lower than the one achieved by DAMA. 
\noindent The background model described in this paper will serve as a comparison tool for the interpretation of the measured background and as a guiding tool throughout the design phase of the full scale experiment.

\section*{Acknowledgements}
The SABRE program is supported by funding from INFN (Italy), NSF (USA), 
and ARC (Australia, grants LE170100162, LE16010080, DP170101675, LP150100075). F.~Froborg has received funding from the European Union's Horizon 2020 
research and innovation programme under the Marie Sklodowska-Curie grant agreement No 703650.
We acknowledge the generous hospitality and constant support of the Laboratori Nazionali del Gran Sasso (Italy).

%
\bibliographystyle{elsarticle-num} 
\bibliography{mybibfile}
\end{document}